\documentclass[sigconf]{acmart}

\usepackage{color}
\usepackage{paralist}
\usepackage{xspace}
\newcommand{\approach}{{\sc DeepMutation}\xspace}
\newcommand{\approachbf}{{\sc \textbf{DeepMutation}}\xspace}
\newcommand{\ie}{\textit{i.e.,}\xspace}
\newcommand{\eg}{\textit{e.g.,}\xspace}
\newcommand{\etc}{\textit{etc.}\xspace}
\newcommand{\etal}{\textit{et al.}\xspace}

\definecolor{pblue}{rgb}{0.13,0.13,1}
\definecolor{pgreen}{rgb}{0,0.5,0}
\definecolor{pred}{rgb}{0.9,0,0}
\definecolor{pgrey}{rgb}{0.46,0.45,0.48}

\usepackage{listings}
\AtBeginDocument{%
  \providecommand\BibTeX{{%
    \normalfont B\kern-0.5em{\scshape i\kern-0.25em b}\kern-0.8em\TeX}}}

\setcopyright{acmcopyright}
\copyrightyear{2020}
\acmYear{2020}
\setcopyright{rightsretained}
\acmConference[ICSE '20 Companion]{42nd International Conference on Software Engineering Companion}{May 23--29, 2020}{Seoul, Republic of Korea}
\acmBooktitle{42nd International Conference on Software Engineering Companion (ICSE '20 Companion), May 23--29, 2020, Seoul, Republic of Korea}\acmDOI{10.1145/3377812.3382146}
\acmISBN{978-1-4503-7122-3/20/05}

\begin{document}

\title{DeepMutation: A Neural Mutation Tool}

\author{Michele Tufano}
\affiliation{%
  \institution{Microsoft}
  \city{Redmond, WA}
  \country{USA}
}
%\email{michele.tufano@microsoft.com}

\author{Jason Kimko, Shiya Wang}
\affiliation{%
  \institution{William \& Mary}
  \city{Williamsburg, VA}
  \country{USA}
}
%\email{{jskimko,swang22}@email.wm.edu}

\begin{comment}
\author{Jason Kimko}
\affiliation{%
  \institution{William \& Mary}
  \city{Williamsburg, VA}
  \country{USA}
}
\email{jskimko@email.wm.edu}

\author{Shiya Wang}
\affiliation{%
  \institution{William \& Mary}
  \city{Williamsburg, VA}
  \country{USA}
}
\email{swang22@email.wm.edu}
\end{comment}

\author{Cody Watson}
\affiliation{%
  \institution{Washington and Lee University}
  \city{Lexington, VA}
  \country{USA}
}
%\email{cwatson@wlu.edu}

\author{Gabriele Bavota}
\affiliation{%
  \institution{Universit\`{a} della Svizzera italiana (USI)}
  \city{Lugano}
  \country{Switzerland}
}
%\email{gabriele.bavota@usi.ch}

\author{Massimiliano Di Penta}
\affiliation{%
  \institution{University of Sannio}
  \city{Benevento}
  \country{Italy}
}
%\email{dipenta@usisannio.it}

\author{Denys Poshyvanyk}
\affiliation{%
  \institution{William \& Mary}
  \city{Williamsburg, VA}
  \country{USA}
}
%\email{denys@cs.wm.edu}

%%
%% By default, the full list of authors will be used in the page
%% headers. Often, this list is too long, and will overlap
%% other information printed in the page headers. This command allows
%% the author to define a more concise list
%% of authors' names for this purpose.
\renewcommand{\shortauthors}{}

\begin{abstract}
Mutation testing can be used to assess the fault-detection capabilities of a given test suite. To this aim, two characteristics of mutation testing frameworks are of paramount importance: (i) they should generate mutants that are representative of real faults; and (ii) they should provide a  complete tool chain able to automatically generate, inject, and test the mutants. To address the first point, we recently proposed an approach using a Recurrent Neural Network  Encoder-Decoder architecture to learn mutants from $\sim$787k faults mined from real programs. The empirical evaluation of this approach confirmed its ability to generate mutants representative of real faults. In this paper, we address the second point, presenting \approach, a tool wrapping our deep learning model into a fully automated tool chain able to generate, inject, and test mutants learned from real faults.  
\textbf{Video:} \url{https://sites.google.com/view/learning-mutation/deepmutation}
\end{abstract}

%%
%% The code below is generated by the tool at http://dl.acm.org/ccs.cfm.
%% Please copy and paste the code instead of the example below.
%%



\maketitle

\section{Introduction}
\label{sec:intro}
The goal of mutation testing is to inject artificial faults into a program~\cite{Hamlet:TSE,DeMillo:Computer} to simulate defects. Mutation testing has multiple applications, including guiding developers to write a test suite (which discovers as many artificial faults as possible), or driving automatic test data generation \cite{FraserA13}. Also, mutants can be used to assess the effectiveness of a test suite or the extent to which a testing strategy discovers additional faults with respect to another~\cite{BriandPL04}.

A crucial issue in mutation testing is the representativeness of mutation operators --- \ie types of artificial faults to be seeded --- with respect to real faults. Empirical research in this area highlights advantages of mutation testing (carefully-selected mutants can be as effective as real faults
\cite{AndrewsBLN06,DaranT96}), but also disadvantages (mutants might underestimate a test suite effectiveness \cite{AndrewsBLN06}). As a result, literature suggests to carefully devise mutant taxonomies \cite{Just:FSE14}, especially to reflect domain-specific characteristics of the software. While past research has proposed mutants for specific domains
\cite{DBLP:conf/sigsoft/JabbarvandM17, wang2015mualloy, 7107450, Offutt:2001:MUO:571305.571314, VasquezBTMPVBP17,deepmutationplusplus-ase2019,MuSC-ase2019}, such a task has been recognized to be effort-intensive.

We have proposed a Neural Machine Translation (NMT) approach to automatically learn mutants from a large corpus of existing bug fixes  mined from software repositories \cite{icsme2019-mutation}. After changes related to bug-fixes have been extracted using an AST-based differencing tool \cite{DBLP:conf/kbse/FalleriMBMM14}, they are grouped into clusters, where each cluster represents a common, recurring set of changes. After that, a Recurrent Neural Network (RNN) Encoder-Decoder architecture \cite{kalchbrenner-blunsom:2013:EMNLP, DBLP:journals/corr/SutskeverVL14, DBLP:journals/corr/ChoMGBSB14}  first learns a model from the corpora of bug-fixing changes, and then applies it to the unseen code.
Key advantages of the proposed approach --- differently from previous work by Brown \etal, which extracts all possible mutations from fixing changes ---
are its capability to determine where and how to mutate source code, as well as its capability to introduce new (unseen) literals and identifiers in the mutated code.

This paper describes \approach, the tool that implements the NMT-based mutation approach we have previously proposed \cite{icsme2019-mutation}. Overall, the major contributions of the paper are:

\begin{compactitem}
    \item The design and release of \approach, the first mutation tool based on an NMT model;
    \item The tool   is available as open source on GitHub \cite{githubAppendix};
    \item The summary of the evaluation we performed for the mutation approach behind \approach \cite{icsme2019-mutation} and a study showing the applicability of the tool on real systems.
\end{compactitem}

\section{Architecture}
\label{sec:approach}
\begin{figure}
    \centering
    \includegraphics[width=0.48\textwidth]{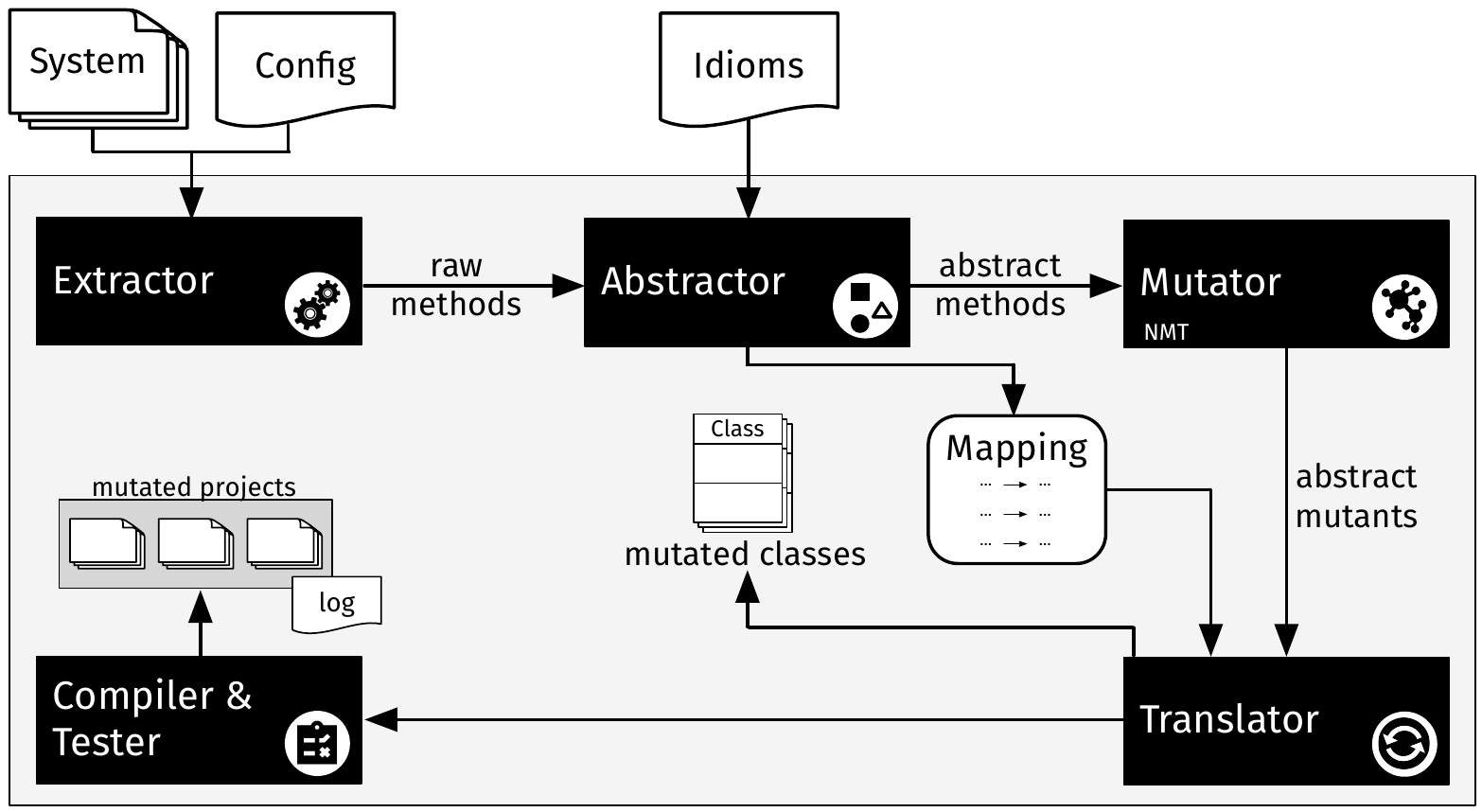}
    \caption{\vspace{-0.4cm}Overview of \approachbf}
    \label{fig:overview}
    \vspace{-0.4cm}
\end{figure}

Figure \ref{fig:overview} provides an overview of the architecture of \approach. Given as input a configuration file and a software system to mutate, \approach~ begins by invoking an \textit{Extractor}, which builds a meta-model of the project and extracts the methods for which mutants will be generated. Each method is then transformed by an \textit{Abstractor}, which creates an abstracted representation of the method. The abstracted method is fed to the \textit{Mutator} which, using a trained NMT model, generates $n$ (user-specified) different abstract mutations. Subsequently, the abstract mutant is translated into source code by the \textit{Translator}. Finally, the \textit{Tester} compiles and tests the generated mutants.

\subsection{Extractor}
\approach~ starts by reading the user-specified settings from a configuration file, storing the paths to the software under test, output folders, and other mutation preferences (details on how to use our tool, starting with the setting of the configuration file, are available in our GitHub repository \cite{githubAppendix}).

The first component to execute is the \textit{Extractor}, which starts by building a meta-model of the software project using Spoon \cite{Pawlak:2016:SLI:3035074.3035076}. Spoon is an open-source library capable of building a well-designed Abstract Syntax Tree (AST), on top of which it provides powerful analysis and transformation APIs. In our case, we primarily use this model to query program elements (\ie specifically methods, but we plan, as future work, to allow \approach~ to work on different granularities) from the software project to analyze, filter, and extract their source code.

Specifically, the \textit{Extractor} queries all the concrete methods in the software system, disregarding abstract or interface methods. It can optionally discard getters/setters methods. %, if the user specified so in the configuration file.
Additionally, the \textit{Extractor} can select only the methods of interest for the user, by matching the signature of the methods with those listed in a user-provided file.
The \textit{Extractor} outputs a list of methods $M$ from the system which pass the filtering criteria.

\subsection{Abstractor}
The main goal of the \textit{Abstractor} is to transform the source code of the methods selected by the \textit{Extractor} into an abstract representation more suitable for our NMT-based \textit{Mutator}. This abstract form is a high-level representation, which reduces the number of tokens used to represent the method (\ie vocabulary size) yet retains all the method's syntactic and semantic information. It contains the following types of tokens:

\textit{Java keywords and separators}: the \textit{Abstractor} keeps all the keywords and separators in order to guarantee syntactic correctness.\smallskip

\textit{Idioms}: frequent identifiers and literals (\eg \texttt{size}, \texttt{index}, \texttt{0}, \texttt{1},  \etc), which we refer to as \textit{idioms}, are retained in the abstract representation, since they provide meaningful semantic information.\smallskip

\textit{Typified IDs}: the \textit{Abstractor} replaces any other tokens (not idioms nor keywords/separators) with IDs, which represents the role and type of the identifier/literal replaced.\smallskip

Specifically, for each method $m \in M$ the \textit{Abstractor} starts from a canonical form of its source code, where any comments or annotations are removed and types are expressed in a fully qualified fashion. Next, a Java Lexer (based on ANTLR \cite{Parr:2011:LFA:1993498.1993548, Parr:2013:DAR:2501720}) reads the source code tokenizing it into a stream of tokens. The tokenized stream is then fed into a Java Parser \cite{javaparser}, which discerns the role of each identifier (\ie whether it represents a variable, method, or type name) and the type of literals (\eg string, char, int, \etc).

The \textit{Abstractor} generates and substitutes a unique ID for each identifier/literal within the tokenized stream. If an identifier or literal appears multiple times in the stream, it will be replaced with the same ID. IDs are represented using the following format: \texttt{<TYPE>\_\#}, where \texttt{<TYPE>} refers to the role of the identifier (\ie \texttt{VAR}, \texttt{TYPE}, \texttt{METHOD}) or type of literal (\ie \texttt{STRING}, \texttt{CHAR}, \texttt{INT}, \texttt{FLOAT}), and $\#$ is a numerical ID. These Typified IDs are assigned to identifiers and literals in a sequential and positional fashion. Thus, the first string literal found will be assigned the ID of \texttt{STRING\_0}, likewise the second string literal will receive the ID of \texttt{STRING\_1}. 

This process continues for all identifiers and literals in the given method $m$, except for those recognized as idioms, that will be kept in the abstract version $m_a$ with their original token. The list of idioms can be specified in a text file by the user, while we provide a default list of 300 idioms mined from thousands of GitHub repositories.

Here we provide an example of how the \textit{Abstractor} would transform the following method:

\begin{lstlisting}[frame=single]
public Integer getMinElement(List myList) {
    if(myList.size() >= 1) {
        return ListManager.min(myList);
    }   
    return null;
}
\end{lstlisting}
into the following abstract stream of tokens, where we highlighted java keywords in blue, idioms in green, and typified IDs in red:
\begin{lstlisting}[frame=single, breakautoindent=false, breakindent=0pt, breaklines=true]
public ~TYPE_1~ ~METHOD_1~ ( <List> ~VAR_1~ ) { if ( ~VAR_1~ . <size> ( ) >= <1> ) { return ~TYPE_2~ . <min> ( ~VAR_1~ ) ; } return null ; }
\end{lstlisting}

In conclusion, for each method $m \in M$ the \textit{Abstractor} generates an abstracted version $m_a$ and a mapping file $Map$ where the mapping of identifiers/literals with their corresponding IDs is stored.

\subsection{Mutator}
The \textit{Mutator} takes as input an abstracted method $m_a$ and generates $k$ different mutants $m^{\prime}_{a1} \dots m^{\prime}_{ak}$, where $k$ is a user-specified parameter defining the number of mutants to generate for each method. The \textit{Mutator} relies on a NMT model that was trained on real bug-fixes mined from thousands of GitHub repositories. 
Given a bug-fix $bf = (m_b, m_f)$ which modifies a buggy method $m_b$ into a fixed method $m_f$, and their abstracted representation $bfa = (m_{ab}, m_{af})$, the NMT model is trained to learn the translation $m_{af} \rightarrow m_{ab}$, that is, learning to generate a buggy method given a method without a bug. In the next paragraphs, we provide technical details behind the NMT model and the approach used to generate multiple mutants.

\subsubsection{NMT Model}
Our model is based on an RNN Encoder-Decoder architecture, commonly adopted in Machine Translation \cite{kalchbrenner-blunsom:2013:EMNLP, DBLP:journals/corr/SutskeverVL14, DBLP:journals/corr/ChoMGBSB14}. This model comprises two major components: an RNN Encoder, which \textit{encodes} a sequence of terms \boldmath{$x$} into a vector representation, and a RNN Decoder, which \textit{decodes} the representation into another sequence of terms $y$. The model learns a conditional distribution over a (output) sequence conditioned on another (input) sequence of terms: \unboldmath$P(y_1,.., y_m | x_1,.., x_n)$, where $n$ and $m$ may differ. In our case, given an input sequence $\mathbf{x} = m_{af} = (x_1,.., x_n)$ and a target sequence $\mathbf{y} = m_{ab} = (y_1,.., y_m)$, the model is trained to learn the conditional distribution:  $P(m_{ab} | m_{af}) = P(y_1,.., y_m | x_1,.., x_n)$, where $x_i$ and $y_j$ are abstracted source tokens: Java keywords, separators, IDs, and idioms.
The Encoder takes as input a sequence $\mathbf{x} = (x_1,.., x_n)$ and produces a sequence of states $\mathbf{h} = (h_1,.., h_n)$. We rely on a bi-directional RNN Encoder \cite{DBLP:journals/corr/BahdanauCB14} which is formed by a backward and forward RNNs, which are able to create representations taking into account both past and future inputs \cite{DBLP:journals/corr/BritzGLL17}. That is, each state $h_i$ represents the concatenation of the states produced by the two RNNs reading the sequence in a forward and backward fashion: $h_i = [ \overrightarrow{h_i}; \overleftarrow{h_i}] $. The RNN Decoder predicts the probability of a target sequence $\mathbf{y} = (y_1,.., y_m)$ given $\mathbf{h}$. Specifically, the probability of each output term $y_i$ is computed based on: (i) the recurrent state $s_i$ in the Decoder; (ii) the previous $i-1$ terms $(y_1,.., y_{i-1})$; and (iii) a context vector $c_i$.  The latter constitutes the attention mechanism. The vector $c_i$ is computed as a weighted average of the states in $\mathbf{h}$, as follows: $c_i = \sum_{t=1}^{n}{a_{it}h_t}$ where the weights $a_{it}$ allow the model to pay more \textit{attention} to different parts of the input sequence. Specifically, the weight $a_{it}$ defines how much the term $x_i$ should be taken into account when predicting the target term $y_t$. The entire model is trained end-to-end (Encoder and Decoder jointly) by minimizing the negative log likelihood of the target terms, using stochastic gradient descent.

\subsubsection{Generating Multiple Mutants via Beam Search}
The main intuition behind Beam Search decoding is that rather than predicting at each time step the token with the best probability, the decoding process keeps track of $k$ hypotheses (with $k$ being the beam size or width). Formally, let $\mathcal{H}_t$ be the set of $k$ hypotheses decoded till time step $t$:
$
\mathcal{H}_t = \{ (\tilde{y}^{1}_1, \dots , \tilde{y}^{1}_t) , (\tilde{y}^{2}_1, \dots , \tilde{y}^{2}_t) , \dots , (\tilde{y}^{k}_1, \dots , \tilde{y}^{k}_t)   \}
$

At the next time step $t+1$, for each hypothesis there will be $|V|$ possible $y_{t+1}$ terms ($V$ being the vocabulary), for a total of $k \cdot |V|$ possible hypotheses:
$
\mathcal{C}_{t+1} = \bigcup\limits_{i=1}^{k}  \{ (\tilde{y}^{i}_1, \dots , \tilde{y}^{i}_t, v_1) , \dots , (\tilde{y}^{i}_1, \dots , \tilde{y}^{i}_t, v_{|V|})   \}
$

From these candidate sets, the decoding process keeps the $k$ sequences with the highest probability. The process continues until each hypothesis reaches the special token representing the end of a sequence. We consider these $k$ final sentences as candidate mutants for the given method $m$. When $k=1$, Beam Search decoding coincides with the greedy strategy.

%\todo{Not sure if we want to have a figure for Beam search}
%Fig. \ref{fig:beamsearch} shows an example of the Beam Search decoding strategy with $k=3$. Given the $m_a$ code as input (top-left), the Beam Search starts by generating the top-3 candidates for the first term (\ie \texttt{public}, \texttt{void}, \texttt{private}). At the next time step, the beam search expands each current hyphothesis and finds that the top-3 most likely are those following the node \texttt{public}. Therefore, the other two branches (\ie \texttt{void}, \texttt{private}) are pruned (\ie red nodes). The search continues till each hypothesis reaches the \texttt{<eos>} (End Of Sequence) symbol. Note that each hypothesis could reach the end at different time steps.

\subsection{Translator}
The \textit{Translator} takes as input an abstract mutant $m^{\prime}_a$ generated by the \textit{Mutator}, translates it in real source code to generate an actual mutant $m^\prime$, which is then injected in the subject system by replacing the existing method $m$. The \textit{Translator} relies on the mapping $Map$ to replace back all the original identifiers/literals in place of the typified IDs introduced by the \textit{Abstractor} and modified by the \textit{Mutator}. While the \textit{Mutator} has been trained to re-use only IDs available in the input method or idioms in the vocabulary, it is still possible that the \textit{Mutator} might introduce an ID which is not present in the original mapping $Map$ (\ie \texttt{<TYPE>\_\#} $\notin Map$). 

These cases are detected by the \textit{Translator}, which will perform a series of best-effort attempts to generate real source code for the mutant. In particular, if the missing IDs refer to literals (\ie \texttt{STRING}, \texttt{CHAR}, \texttt{INT}, \texttt{FLOAT}), it will generate a new literal (of the specified type) making sure that is not already defined in the original method $m$ (since it would have been referred already with an ID). Conversely, if the missing IDs refer to variables or method calls, the \textit{Translator} will discard the current mutant and select the next one.

Once all the IDs in $m^{\prime}_a$ are replaced, we obtain the concrete code of the mutant $m^{\prime}$. Next, the code is automatically formatted and indented. Finally, \approach~ replaces the original method $m$ with the newly generated and translated mutant $m^{\prime}$.

%At this stage, \approach~ stores only copies of the Java class files where the mutants have been injected, not entire copies of the software projects. This minimizes the needed storage. In the next step, \approach~ replaces these classes on-the-fly within the subject system to compile and test them. 

\subsection{Compiler \& Tester}
%\MICHELE{Not sure if we should call it just Compiler, Tester, or both. We did not Test the mutants, but we have the capabilities to do so (the user can specify the commands and we run the tests)} \GABRIELE{Let's keep both}

\approach~ allows the user to specify custom commands for compilation and testing of the mutants in the configuration file. This enables \approach~ to compile and test every mutant generated and output a log reporting the results of these steps.

%\subsection{Installation \& Run}
%\label{sub:installation}
%We conclude this section on the architecture of \approach~ with the instructions on how to install and run this tool.

%The following command guides the user through the installation of dependencies and set-up of the environment:
%\begin{lstlisting}
%./setup_env.sh
%\end{lstlisting}

%Next, to build \approach, run this \texttt{ant} command:
%\begin{lstlisting}
%ant dist
%\end{lstlisting}

%Finally, run \approach~ as a jar file providing the configuration file:
%\begin{lstlisting}
%java -jar DeepMutation.jar config_file
%\end{lstlisting}

%More info is available in our GitHub repository \cite{githubAppendix}.

\section{Evaluation}
\label{sec:evaluation}
The \textit{goal} of this evaluation is to assess the performance of the end-to-end mutation infrastructure offered by \approach. A more comprehensive evaluation of the quality of the mutants generated by the NMT model can be found in \cite{icsme2019-mutation}.

We execute the entire pipeline of \approach~ on four projects from the Defects4j \cite{Just:ISSTA14} dataset. Specifically, we select the first (fixed) revision of the following projects: Chart, Lang, Math, Time. 

\subsection{Settings}
As \approach~ core NMT model we rely on a pre-trained model from our previous work \cite{icsme2019-mutation}. This model was trained on a large, diverse set of bug-fixes mined from thousands of GitHub repositories. For this evaluation, we do not fine-tune nor re-train the NMT model on the Defects4J codebase. This allows us to demonstrate the capabilities of the model on a novel, previously unseen codebase and provide insights on the transfer learning opportunities.

We set the beam size equal to one, thus allowing the model to generate only a single mutant for each method extracted from the project. Note that \approach~ supports the generation of many different mutants for each method.

The maximum method size is set to 50 tokens for the purpose of this evaluation. Thus, we disregard methods larger than 50 tokens after the extraction process and before the abstraction. \approach~ supports the mutation of methods larger than 50 tokens.

\subsection{Mutation Performances}
Table \ref{tab:results} shows the results of our preliminary evaluation in terms of the number of methods extracted and mutated, throughout the different phases of \approach. The \textit{Extractor} identified between 3k-7k methods in each project. After this stage, we automatically discard getters/setters as well as methods longer than a predefined threshold. Between 1.1k-2.8k methods are abstracted and mutated for each project. Next, the \textit{Translator} was able to correctly translate abstract mutants generated by the \textit{Mutator} in concrete source code in 82\%-97\% of the cases. Finally, \approach~ creates a mutated version of the project for each mutant, by injecting the mutant in a copy of the project and attempting to compile the entire mutated project. \approach was able to successfully compile 35\%-67\% of the mutated projects. 
We claim this is a satisfactory level of compilability, although we aim to improve it, in future work, by
 dynamically replacing non-compilable mutants with the next probable mutant obtained via beam search.

\subsection{Timing Performances}
Table \ref{tab:time} provides details on the timing performance of \approach. We report the total mutation time for all the mutants for a given project, the time spent specifically by the compiler, and the thread load during these experiments. The results show that \approach~ can mutate thousands of methods (see Table \ref{tab:results}) in a reasonable amount of time. Specifically, if we exclude the time spent compiling the projects, \approach~ takes between 5-12min to extract, abstract, mutate, and translate the mutants.

\begin{table}[]
\centering
\caption{Preliminary Evaluation Results\vspace{-0.4cm}}
\label{tab:results}
\resizebox{\columnwidth}{!}{%
\begin{tabular}{lrrrr}
\toprule
Project     & Extractor & Abstractor/Mutator    & Translator        & Compiler \\ \midrule
Chart.1     & 3,383     & 1,134                 & 1,103 (97.27\%)   & 741 (67.18\%)\\ 
Lang.1      & 4,350     & 1,607                 & 1,322 (82.27\%)   & 585 (44.25\%)\\ 
Math.1      & 8,988     & 2,810                 & 2,416 (85.98\%)   & 1,397 (57.82\%)\\ 
Time.1      & 7,047     & 2,028                 & 1,941 (95.71\%)   & 687 (35.39\%)\\
\bottomrule
\end{tabular}
}
\end{table}

\begin{table}[]
\centering
\caption{Timing Performances\vspace{-0.4cm}}
\label{tab:time}
\resizebox{\columnwidth}{!}{%
\begin{tabular}{lrrr}
\toprule
Project     & Thread Load   & Compiler Time     & Total Mutation Time \\ \midrule
Chart.1     & 34/35         & 4min              &9min   \\
Lang.1      & 41/42         & 5min              &12min  \\ 
Math.1      & 75/76         & 15min             &27min  \\ 
Time.1      & 60/61         & 16min             &24min  \\
\bottomrule
\end{tabular}
}
\vspace{-0.3cm}
\end{table}

\section{Conclusions}
\label{sec:conclusion}
We presented \approach, a mutation infrastructure featuring an NMT model trained to generate mutants that resemble real bugs \cite{icsme2019-mutation}. We described the infrastructure,  able to extract, abstract, and mutate methods from a given project, as well as automatically compile and test the generated mutants. We provided an evaluation of the infrastructure by generating mutants for four projects.% In future, we plan to keep improving \approach. %, in particular the compilation rate.%, and perform a more comprehensive test of the generated mutants.

\section{Acknowledgment}
\label{sec:acknowledgment}
This work is supported in part by the NSF CCF-1927679, CCF-1815186, CNS-1815336 grants. Bavota acknowledges the financial support of the Swiss National Science Foundation for the project CCQR (SNF Project No. 175513).

\bibliographystyle{ACM-Reference-Format}
\bibliography{ms}

\end{document}